\documentclass[letterpaper, onecolumn, 11pt]{article}
\usepackage[left=2cm,right=2cm,top=2cm,bottom=2cm]{geometry}

\usepackage[utf8x]{inputenc}
\usepackage{url}
\usepackage{microtype}
\usepackage{graphicx}
\usepackage{booktabs}
\usepackage{tabulary}
\usepackage{subfigure}

\usepackage{setspace}

\usepackage[unicode=true]{hyperref}

\hypersetup{breaklinks=true,
            bookmarks=true,
            pdfauthor={},
            pdftitle={},
            colorlinks=true,
            urlcolor=blue,
            linkcolor=magenta,
            pdfborder={0 0 0}}

\title{Evolution of the Chilean Web: A Larger Study}

\author{Eduardo Graells-Garrido and Ricardo Baeza-Yates \\
Center for Web Research \\
Dept. of Computer Science \\
Universidad de Chile \\
Santiago, Chile}

\date{}

\begin{document}

\maketitle
\thispagestyle{empty}

\begin{abstract}
In this paper we extend our previous and only study on the dynamics of the Chilean Web.
This new study doubles the time period and to the best of our knowledge is the only study
of its type known about any country in the Web. The new results corroborate the trends found 
before, in particular the exponential growth of the Web, and reinforce the conclusion that 
the Web is more chaotic than we would like. Hence, modeling most Web characteristics is not trivial.
\end{abstract}

\section{Introduction}
\label{sec:introduction}

The Chilean Web was born in year 1993, but the first massive crawling of sites for its characterization was made in year 2000. That first analysis found that some characteristics of the Global Web were also present in the subset represented by the Chilean Web. In the following years more crawls were done by the search engine TodoCL and the Center for Web Research. 

Other countries have been crawled until now. In \cite{baeza2007nationaldomains} a total of 10 national domains are fully studied. All the results obtained come from the analysis of snapshots in a specific time. What happens when we consider a single domain and analyze it through time? Trying to answer this question can be problematic, because of the nature of the Web: not only it's dynamic, also it's very chaotic, as sites appear and disappear constantly. Those who remain also change their content very often, specially nowadays where a great amount of content is displayed in pages and URLs generated on the fly thanks to server technologies. Even considering that chaotic behavior and the fact that every crawl is different, the results obtained every year are coherent, in the sense that they show similar behavior for the different analysis made, so it is natural to try to characterize the dynamics of the Chilean Web though all the years we have collected it.

In \cite{dynamics-web-structure} we focused on the macrostructure of the Chilean Web from 2000 to 2004. In this paper we include data from the following years, up to 2007. To the best of our knowledge, there is no other study on such large timeframe for a subset of the Global Web or, for that matter, of the complete Web. In Section \ref{sec:chilean_web}, we define the Chilean Web and expose its properties, analyzing the documents found per site, site content in \textit{megabytes}, the linkage between sites and PageRank. Section \ref{sec:web_structure} analyzes the evolution of the web structure of the Chilean Web. Finally, our conclusions are shown in Section \ref{sec:conclusions}.

\section{The Chilean Web}
\label{sec:chilean_web}

We define the Chilean Web as all sites under the Top Level Domain (TLD) \texttt{.cl}, plus the sites that are hosted in IP addresses associated to Chile. A site is a collection of pages that share the \textit{hostname} in the URL. Each page can be reached by a unique URL. and each site belongs to a domain, although in this article we only study sites. These concepts can be seen in the following schema:

$$\displaystyle \overbrace{http://\overbrace{\underbrace{www.\underbrace{cwr.\underbrace{cl}_{TLD}}_{Domain}}_{Site}/index.html}^{Document/Page}}^{URL}$$

\begin{table*}[ht]
\begin{center}
\footnotesize
\caption{TodoCL \& Center for Web Research Collections.}
\begin{tabular}{l|cccccccc}
\hline & 2000 & 2001 & 2002 & 2003 & 2004 & 2005 & 2006 & 2007 \\
\hline Pages & $612.910$ & $794.218$ & $2.089.406$ & $2.551.567$ & $3.078.901$ & $2.844.137$ & $7.197.032$ & $9.367.543$ \\
\hline Crawled Sites & $7.497$ & $21.207$ & $38.965$ & $37.132$ & $53.509$ & $45.846$ & $93.583$ & $111.110$ \\
-- New  & $7.497$ & $15.415$ & $23.359$ & $8.978$ & $21.874$ & $7.662$ & $50.484$ & $35.512$ \\
Not Crawled 		& n/a & $2.604$ & $8.205$ & $22.666$ & $28.163$ & $43.488$ & $46.235$ & $64.220$ \\
-- Unknown Status 	& n/a & $1.389$ & $1.417$ & $6.446$  & $3.139$  & $13.154$ & $ 2.441$ & $21.627$ \\
-- Dead 		& n/a & $1.215$ & $6.788$ & $16.220$ & $25.024$ & $30.334$ & $43.794$ & $42.593$ \\
\hline Crawled Domains & $6.280$ & $19.387$ & $35.128$ & $33.241$ & $47.468$ & $36.961$ & $85.484$ & $104.409$ \\
Known Domains & $6.280$ & $20.662$ & $41.436$ & $50.671$ & $69.915$ & $72.163$ & $117.270$ & $150.233$ \\
\hline Content in MBs & $2.300$ & $7.090$ & $4.747$ & $9.242$ & $48.154$ & $10.422$ & $121.902$ & $138.375$ \\
\hline 
\end{tabular}
\label{tab:collections}
\end{center}
\end{table*}

To crawl the Chilean Web, we start with an initial list of sites (\textit{seeds}). From those seeds we download pages, and in those pages we seek new links for download. This cycle is repeated until the crawler reaches one of these conditions: \textit{a)} there are no more pages to crawl (unlikely), \textit{b)} the process reached the maximum number of downloadable pages (probably caused by an underestimation of the size of the crawl) ,\textit{c)} there are more pages in schedule for crawling, but they are dynamic pages or have a greater depth (the number of directories from the root of the site) than the max crawled depth defined in the configuration (this is more likely to happen), and \textit{d)} there is no free space on disk (this could be problematic, but it is a possible scenario).

In 2000, we had an initial list of seeds, therefore the crawl was biased to the sites we knew. Since 2001, NIC Chile (\url{http://www.nic.cl}) thanks to a research agreement has been giving us the list of registered \texttt{.cl} domains, making possible to crawl sites that didn't receive links from anywhere in the Chilean Web. Table \ref{tab:collections} shows the global statistics for each year crawled. The crawled sites are those who had at least one downloadable page, and when a site URL is crawled for the first time, the site is marked as \textit{new}. Also there are non crawled sites, which can belong to two special states: a) \textit{unknown}, or sites that weren't crawled because of technical problems, b) \textit{dead sites}, or sites that didn't have an IP address associated to them. We make a similar distinction in domains, as some of them were crawled and some were in an unknown state, but we cannot determine if a domain is dead because a subdomain still can have an IP address. The content of each crawl is the sum of bytes used in disk for all documents inside the crawl. The metadata associated is not considered, and multimedia files in sites are not downloaded.

In the past, we said the that the Chilean Web grown exponentially in terms of sites \cite{dynamics-web-structure}. Using a least squares fitting on the values of Table \ref{tab:collections}, the growth of the number of sites, pages and content size can be approximated by a exponential growth: $sites_{n} = 1,32 * sites_{n-1}$, $pages_{n} = 1,46 * pages_{n-1}$ and $content_{n} = 1,55 * content_{n-1}$. 

We know that the Global Web resembles a \textit{free scale network} \cite{barabasi2002}, and according to this we have found that the distribution of certain variables in the Chilean Web can be approximated by powerlaws ($k/x^{\theta}$). In this article we highlight the distribution of pages per site, of content size per site, incoming and outcoming links, and PageRank; all these variables have been  approximated by powerlaws using a linear regression on the sampled data. We are mostly interested in the exponent of the powerlaw, because it shows how unequal is the distribution for each variable. Table \ref{tab:powerlaws} shows the estimated powerlaws of the mentioned distributions.

\begin{table*}[htbp]
\begin{center}
\footnotesize
\caption{Approximation of Powerlaws ($\displaystyle k/x^{\theta}$) for characteristics of each crawl. The value of $\theta$ and the range in which the powerlaw fits best the data is shown.}
\begin{tabular}{l|cccccccc|l}
\hline 			& 2000 & 2001 & 2002 & 2003 & 2004 & 2005 & 2006 & 2007 & Range \\
\hline Pages per Site 	& $1,45$ & $1,52$ & $1,68$ & $1,74$ & $1,82$ & $1,81$ & $1,73$ & $1,83$ & $[10,500]$ \\ 
Site Content in MBs 	& --- & $1,80$ & $2,14$ & $1,96$ & $1,67$ & $1,56$ & $1,59$ & $1,81$ & $[1,100]$ \\ 
In-Degree 		& --- & $1,78$ & $1,93$ & $2,19$ & $2,03$ & $2,07$ & $1,94$ & $1,83$ & $[1,100]$ \\ 
Out-Degree 		& --- & $1,71$ & $2,02$ & $1,94$ & $2,19$ & $2,18$ & $1,91$ & $1,84$ & $[1,100]$ \\ 
Sum of PageRank		& --- & --- & --- & --- & $0,95$ & --- & $1,05$ & $1,05$ & $[10^{-7}, 10^{-4}]$ \\
\hline
\end{tabular}
\label{tab:powerlaws}
\end{center}
\end{table*}

\subsection{Documents per site}
\label{sec:documents}

\begin{figure}[htb]
 \centering
 \includegraphics[width=0.7\textwidth]{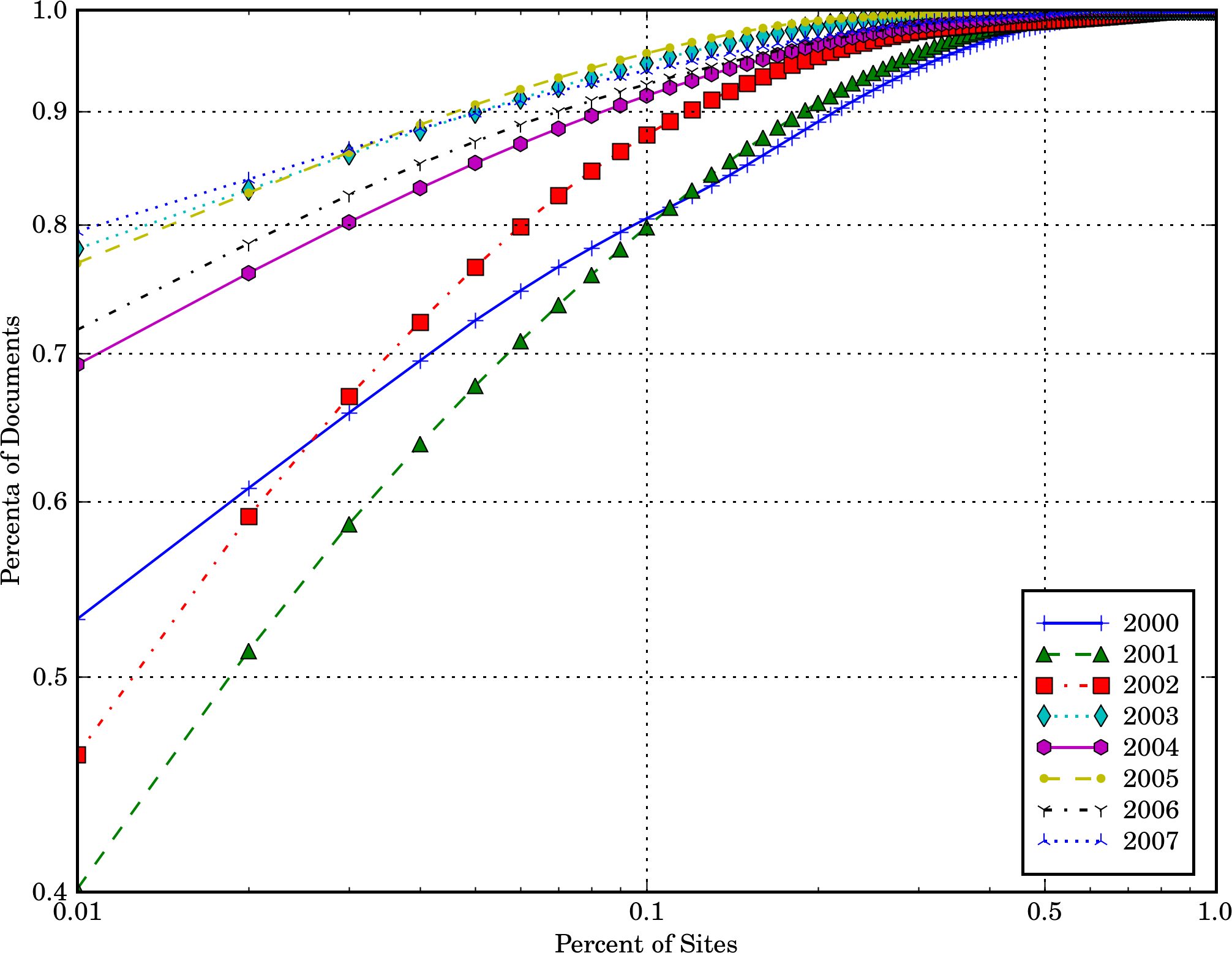}
 \caption{Accumulation of documents per site (2000--2007).}
 \label{fig:acc_documents}
\end{figure}

The distribution of documents per site can be approximated with powerlaws with exponents between $-1,45$ in 2000 and $-1,83$ in 2007, comparable to $-1,45$ in Argentina \cite{tolosa2007characterization}. Those exponents shows that the distribution is getting more unequal each year. This is confirmed by Figure \ref{fig:acc_documents}, which shows the fraction of sites versus the fraction of total documents per year. Every year a fewer fraction of sites contains a bigger fraction of the total documents. 

It should be noted that the distribution is highly chaotic: the sites with most pages in 2007 were not in the list of sites with most documents in 2006, the sites with most pages in 2006 weren't in the list for 2005, and so on. There are reasons for this chaotic behavior: a) the birth and death of sites, b) the proliferation of dynamic pages, mainly because of \textit{Content Management Systems} (CMSs) such as product catalogs, forums and blogs, c) the disguise of dynamic pages as static pages because of friendly URL techniques, such as \textit{URL Rewriting}, which make a dynamic URL look like a static document. This makes the depth restriction and dynamic page detection harder to apply, and thus, there are more crawled pages per site than it should be. 

Regarding the sites with less pages, every year, an average of $39\%$ of the crawled sites have only one page, or at least one crawled page. In the second case, the other pages could not be crawled because of browser-dependant navigation (like Javascript/Flash/Java menus or AJAX actions). In 2007, we estimated that a $64,91\%$ of one page sites used browser-dependant technology. Table \ref{tab:1_page_sites} shows the total sites with one page crawled per year.
 
\begin{table}[htbp]
\begin{center}
\footnotesize
\caption{Number and percent of sites with only one crawled page.}
\begin{tabular}{l|cc}
\hline Year & 1 Page Sites & \% of Crawled Sites\\ 
\hline 2000 & $3.029$  & $40,40$ \\ 
2001 & $9.193$  & $43,35$ \\ 
2002 & $12.521$ & $31,44$ \\ 
2003 & $13.699$ & $36,89$ \\ 
2004 & $21.450$ & $40,09$ \\ 
2005 & $20.411$ & $44,52$ \\ 
2006 & $36.654$ & $39,17$ \\ 
2007 & $48.103$ & $43,29$  \\
\hline
\end{tabular}
\label{tab:1_page_sites}
\end{center}
\end{table}

\subsection{Site content in megabytes}
\label{sec:size}

The powerlaws for site content are not as stable as the ones for documents per site, which clearly had a tendency. This can be explained as the distribution of site content is as chaotic as the distribution of documents, because it is affected by the same anomalies, but it also depends on additional crawler policies (we established a limit of 100 KB per page, which is sufficient for most pages) and the text content of sites. 

\begin{figure}[htb]
 \centering
 \includegraphics[width=0.7\textwidth]{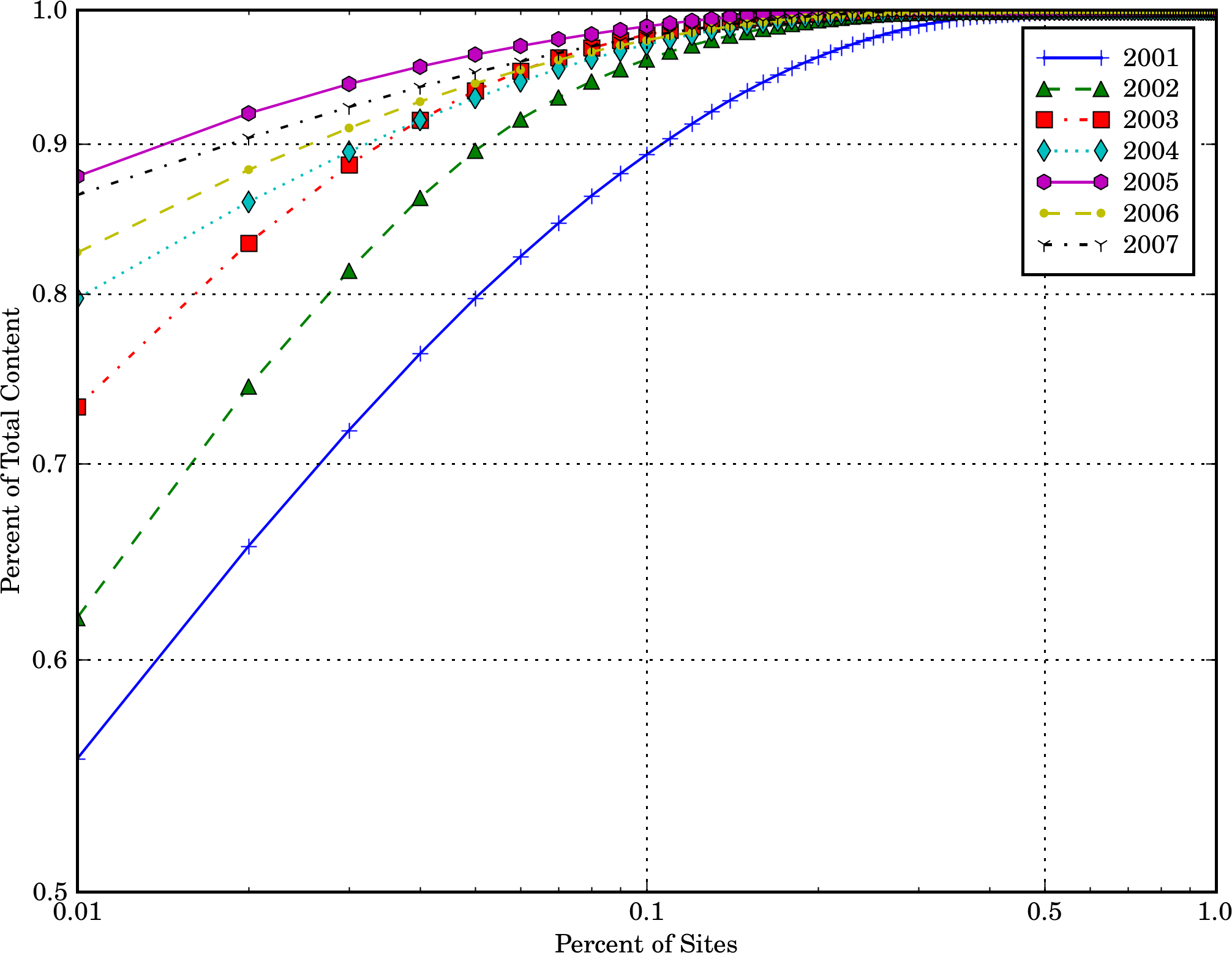}
 \caption{Accumulation of site content in MBs (2001--2007).}
 \label{fig:acc_size}
\end{figure}

Figure \ref{fig:acc_size} shows the accumulation of site content through the years. In 2007, $1\%$ of the crawled sites had $87\%$ of the total content in MBs, although in 2005 the distribution was more biased. The powerlaws have exponents between $-2,14$ and $-1,56$, comparable to $-1,63$ in Korea \cite{baeza04korea}.

\subsection{Links between sites}
\label{sec:links}

\begin{figure*}[tb]
\begin{center}
 \subfigure{\includegraphics[width=0.45\textwidth]{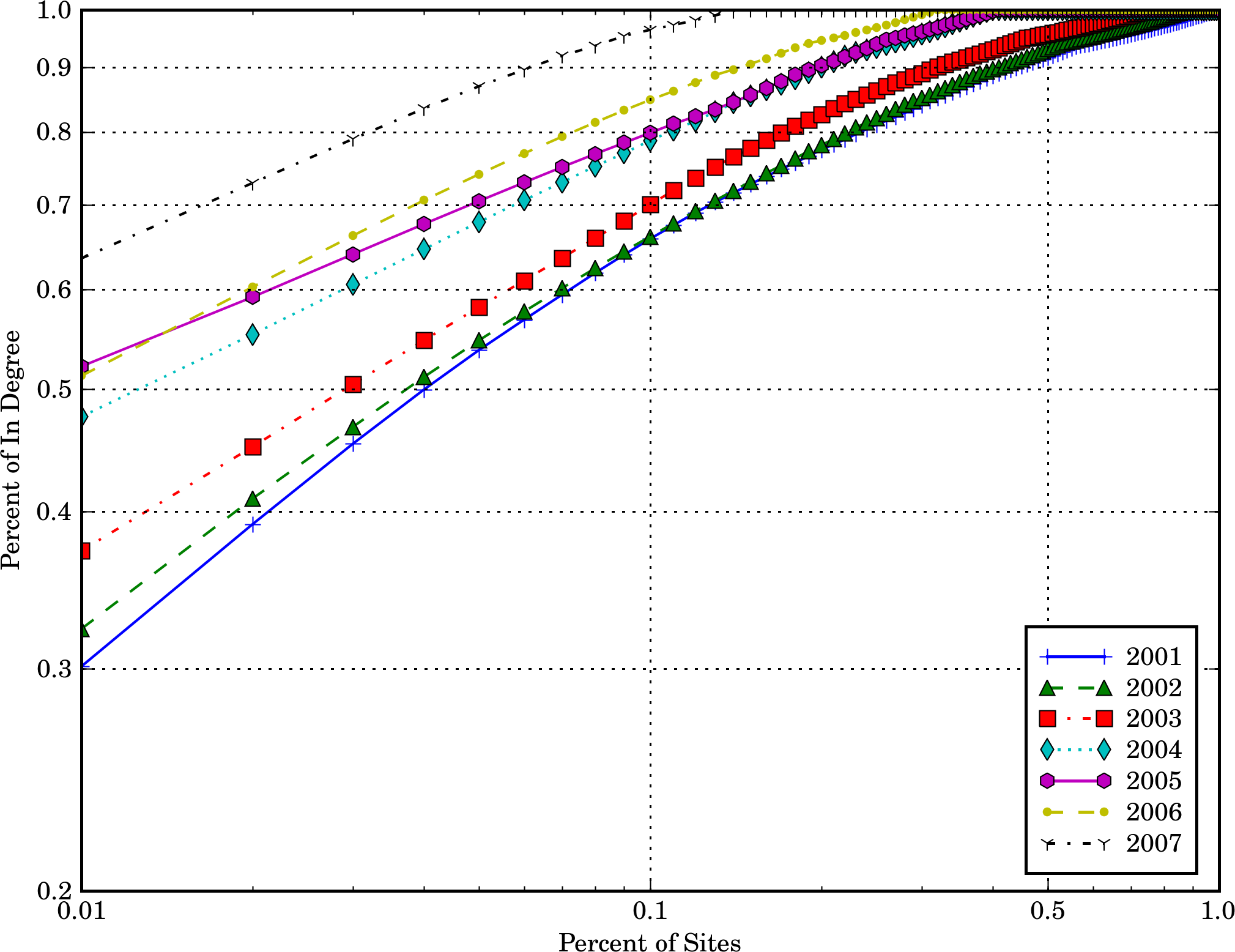}}
 \subfigure{\includegraphics[width=0.45\textwidth]{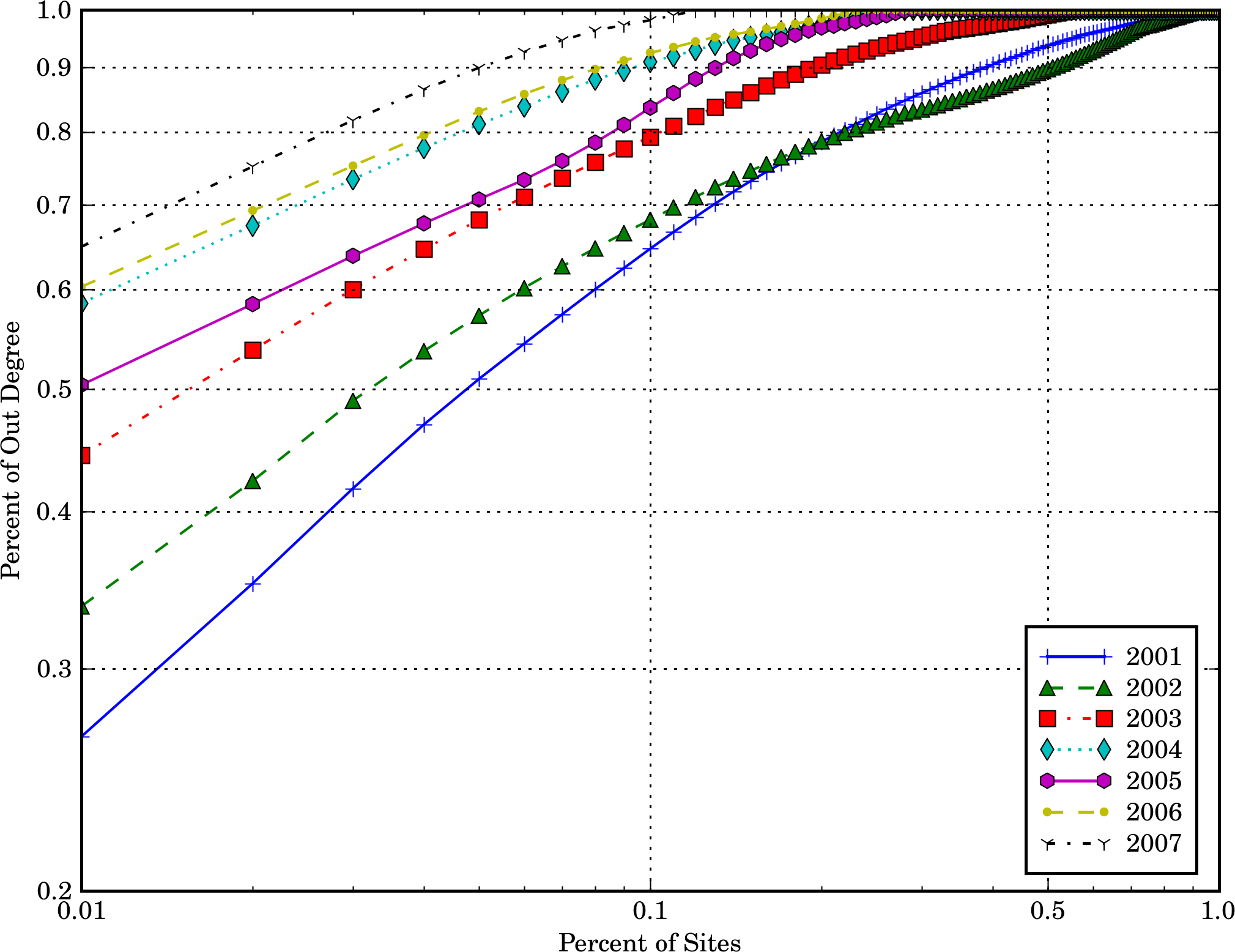}}
 \subfigure{\includegraphics[width=0.45\textwidth]{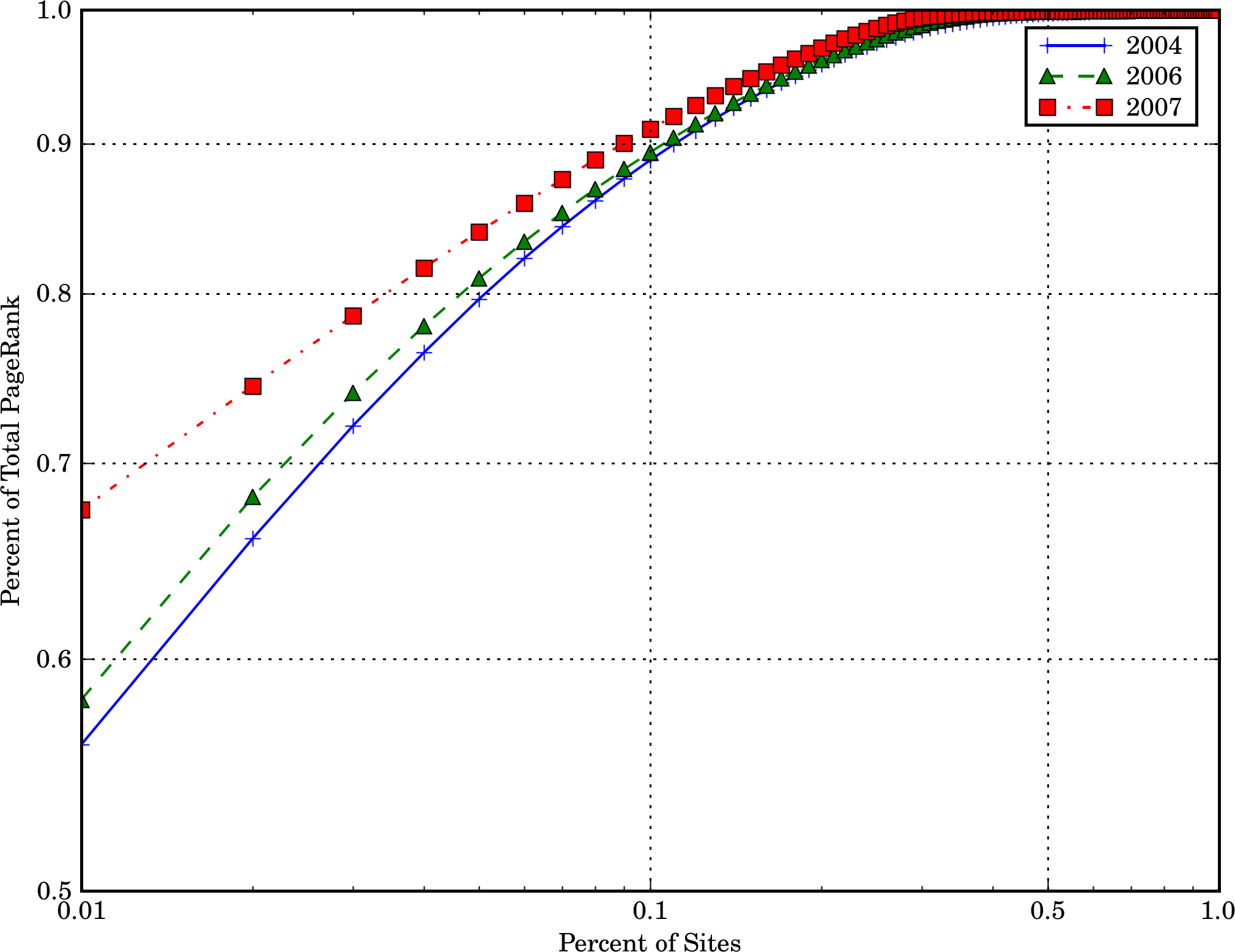}}
\end{center}
\caption{Accumulation of in-degree (Left), out-degree (Center) and Sum of PageRank (Right) for sites in the collections.}
\label{fig:acc_links}
\end{figure*}

A link between two sites represents one or several links between their pages, preserving the direction. This means that if a site $A$ has one or more pages linking to site $B$, then we have a \emph{site link} from $A$ to $B$. The resulting graph is called the \emph{hostgraph} \cite{broder2000graph}, in which the number of incoming links for a site is called \textit{in-degree}, and the number of sites linked from a site is called \textit{out-degree}. Both quantities, in-degree and out-degree, can be approximated by powerlaws for all years: in-degree with exponents between $-2,19$ and $-1,78$ and out-degree with exponents between $-2,19$ and $-1,71$. Surprisingly, both distributions have similar values, while in other national domains they have very different ones. The distributions are comparable to $-1,82$ for in-degree in Spain \cite{baeza06caracteristicas} and $-1,90$ for out-degree in Greece \cite{efthimis04charting}.

Figure \ref{fig:acc_links} shows that the accumulation of in-degree is extremely unequal, and this inequality is increasing every year: in 2007 less than $15\%$ of the crawled sites had the total in-degree of the collection. This can have different interpretations: a) most new sites are not worth linking, b) most new sites do not try or are not interested in getting links, and c) the new sites appear faster than the capability of existing sites to add links to them. 

Figure \ref{fig:acc_links} shows the accumulation of out-degree, which is almost as unequal as in-degree: less than $15\%$ of the crawled sites had the total out-degree in 2007. Out-degree is almost as unequal as in-degree, which is surprising because each site can control its own out-degree. This indicates that every year most of the new sites simply don't want to link other sites.

The Sum of PageRank for a site is defined as the sum of the PageRank associated to each HTML document in the site. We calculated this sum for each site in the years 2004, 2006 and 2007. Because of the nature of the PageRank algorithm, even sites without incoming links have a non-null value. Figure \ref{fig:acc_links} shows the accumulation of PageRank though sites, which is stable across the years: its distribution can be approximated by powerlaws with parameters $-0,95$ (2004) and $-1,05$ (2006 and 2007), comparable to $-1,76$ in Spain \cite{baeza06caracteristicas}.

\subsection{Stable Sites}

Table \ref{tab:stable_sites} shows the 10 most stable sites across the years for each characteristic analyzed. To find these sites, for each year we calculated the fraction of the contribution of each site to the corresponding characteristic. This gave us a normalized contribution for every year. After, we calculated the average contribution for all sites in the period 2000--2007. This produced a sorted list of contributions. Finally, we filtered this list to remove sites that were not crawled in three years or more.

\begin{table*}[htbp]
\begin{center}
\footnotesize
\caption{The most stable sites through 2000--2007 in the distinct characteristics analyzed. Sites from the Government are marked with  \textit{(G)}, Universities and Educational Institutions with \textit{(U)}, Commercial sites with \textit{(C)}, and Newspapers, Magazines and Media with \textit{(M)}. Other kind of sites are marked with \textit{(O)}.}
\begin{tabular}{|l|l|l|l|}
\hline Pages per Site 	& Site Content 		& In-Degree 			& Out-Degree \\ 
\hline (O) \url{www.eclac.cl} 	& (G) \url{www.camara.cl} 	& (G)\url{ www.sii.cl} 		& (C) \url{www.todocl.cl} \\ 
(O) \url{www.cepal.cl} 	& (G) \url{www.bcentral.cl} 	& (G) \url{www.mineduc.cl} 		& (C) \url{www.universia.cl} \\ 
(C) \url{www.tucows.cl} 	& (U) \url{fisica.ciencias.uchile.cl} & (G) \url{www.bcentral.cl} 	& (C) \url{www.servicioweb.cl} \\ 
(M) \url{www.tercera.cl} 	& (C) \url{www.tucows.cl} 	& (U) \url{www.puc.cl} 		& (C) \url{www.123.cl} \\ 
(U) \url{www.udec.cl} 	& (G) \url{www.sofofa.cl} 	& (G) \url{www.corfo.cl} 		& (C) \url{www.yes.cl} \\ 
(U) \url{sunsite.dcc.uchile.cl} & (G) \url{www.mineduc.cl} 	& (U) \url{www.uchile.cl} 		& (U) \url{www.uchile.cl} \\ 
(U) \url{www.uchile.cl} 	& (G) \url{www.conama.cl} 	& (U) \url{www.udec.cl} 		& (C) \url{www.huellas.cl} \\ 
(C) \url{www.chilnet.cl} 	& (O) \url{www.eclac.cl} 	& (G) \url{www.gobiernodechile.cl} 	& (C) \url{www.nic.cl} \\ 
(G) \url{www.conicyt.cl} 	& (U) \url{www.elo.utfsm.cl} 	& (G) \url{www.meteochile.cl} 	& (C) \url{www.mundopyme.cl} \\ 
(C) \url{www.delcerro.cl} 	& (M) \url{www.quepasa.cl} 	& (G) \url{www.conicyt.cl} 		& (C) \url{www.guia-chile.cl} \\ \hline
\end{tabular}
\label{tab:stable_sites}
\end{center}
\end{table*}

Using the normalized average criteria, the list has a majority of sites from the government and universities in all characteristics except out-degree. The sites with stable out-degree are mostly search engines, directories or portals for diverse subjects. On the other hand, the sites with most in-degree are mostly from government and universities. This list is very stable, as the 10 listed sites are the most linked in almost all years covered by our study. 

As said in Section \ref{sec:documents}, the sites with more pages and content are very unstable. Although the sites in Table \ref{tab:stable_sites} do not appear in the list of bigger sites per year, they indeed have more documents and content than the majority of sites crawled each year. 

\section{Evolution of Web structure}
\label{sec:web_structure}

The structure of a Chilean Web at the site level (hostgraph) is similar to the structure of the Global Web \cite{broder2000graph}, as we found in \cite{baeza01relating}. In the Chilean Web, the hostgraph has one giant strongly connected component (SCC), which defines the following components for the macro-structure of the Web:

\begin{itemize}
 \item[(a)] MAIN, sites in the SCC of the hostgraph. This means that it's possible to reach any site in MAIN from another site in MAIN following links. 
 \item[(b)] IN, sites that can reach MAIN through links but cannot be reached from MAIN.
 \item[(c)] OUT, sites that can be reached from MAIN, but there is no path in links to go back to MAIN.
 \item[(d)] other sites that can be reached from IN (T.IN, where T is an abbreviation for tentacles), sites in paths between IN and OUT (TUNNEL), sites that only reach OUT (T.OUT), and unconnected sites (ISLANDS).
\end{itemize}

In \cite{baeza01relating} we divided MAIN into four subcomponents:

\begin{itemize}
 \item[(a)] MAIN-MAIN, sites that can be reached directly from IN and can reach directly OUT (that is, interconnection sites from IN to OUT). 
 \item[(b)] MAIN-IN, sites that can be reached directly from IN, but are not in MAIN-MAIN.
 \item[(c)] MAIN-OUT, sites that can reach OUT directly, but are not in MAIN-MAIN.
 \item[(d)] MAIN-NORM, sites in MAIN that do not belong to any of the previous subcomponents.
\end{itemize}

\begin{figure}[htb]
 \centering
 \includegraphics[width=0.5\textwidth]{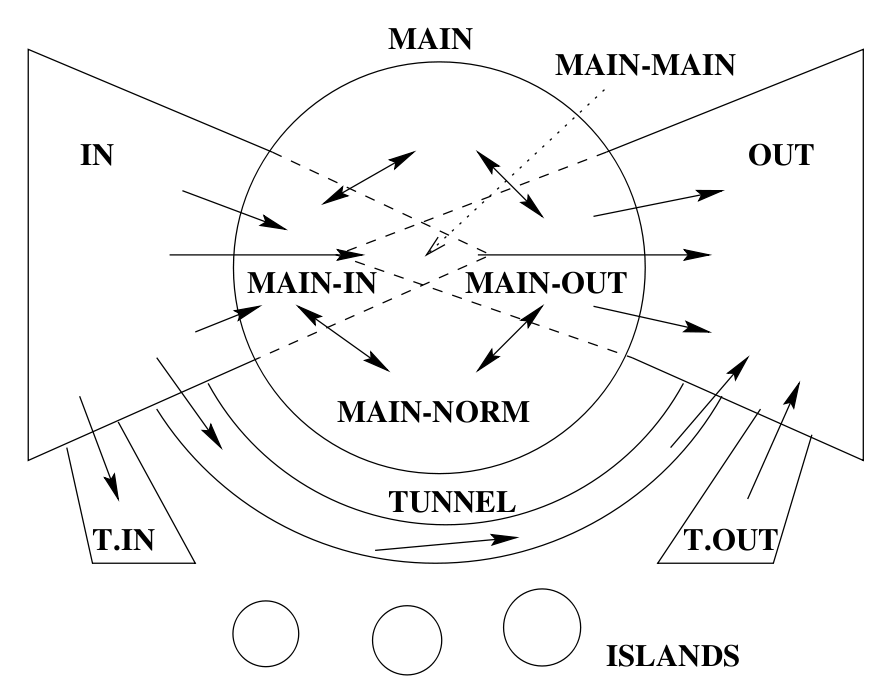}
 \caption{Graphical representation of the Web Structure.}
 \label{fig:structure_sites}
\end{figure}

Given this structure, shown in Figure \ref{fig:structure_sites}, having a good set of seeds it is possible to crawl MAIN (the SCC) and OUT. To find sites in IN, ISLANDS, T.OUT and TUNNEL, the addresses of those sites is needed, because no one inside the national domain points to them. As we have most of the registered domains thanks to NIC Chile, our study has a very large coverage. On the other hand, because any crawling is incomplete (for example, dynamic pages can be unbounded), any Web graph will be incomplete. That means that any analysis of the Web structure will be an approximation. 

\begin{figure}[htb]
 \centering
 \includegraphics[width=0.6\textwidth]{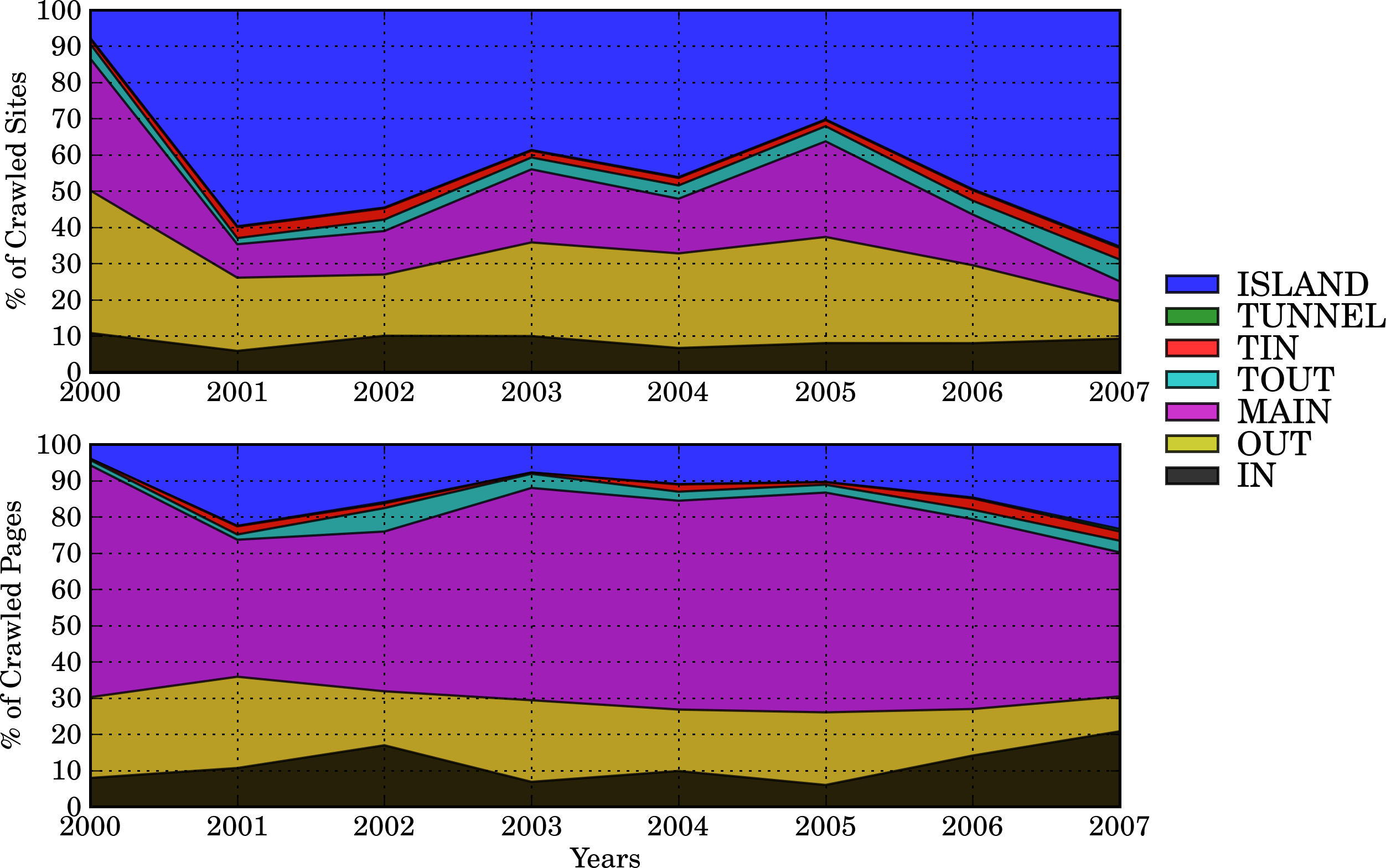}
 \caption{Evolution of the components of the Web Structure in terms of sites (top) and pages (bottom).}
 \label{fig:structure_evolution}
\end{figure}

\begin{figure}[htb]
 \centering
 \includegraphics[width=0.6\textwidth]{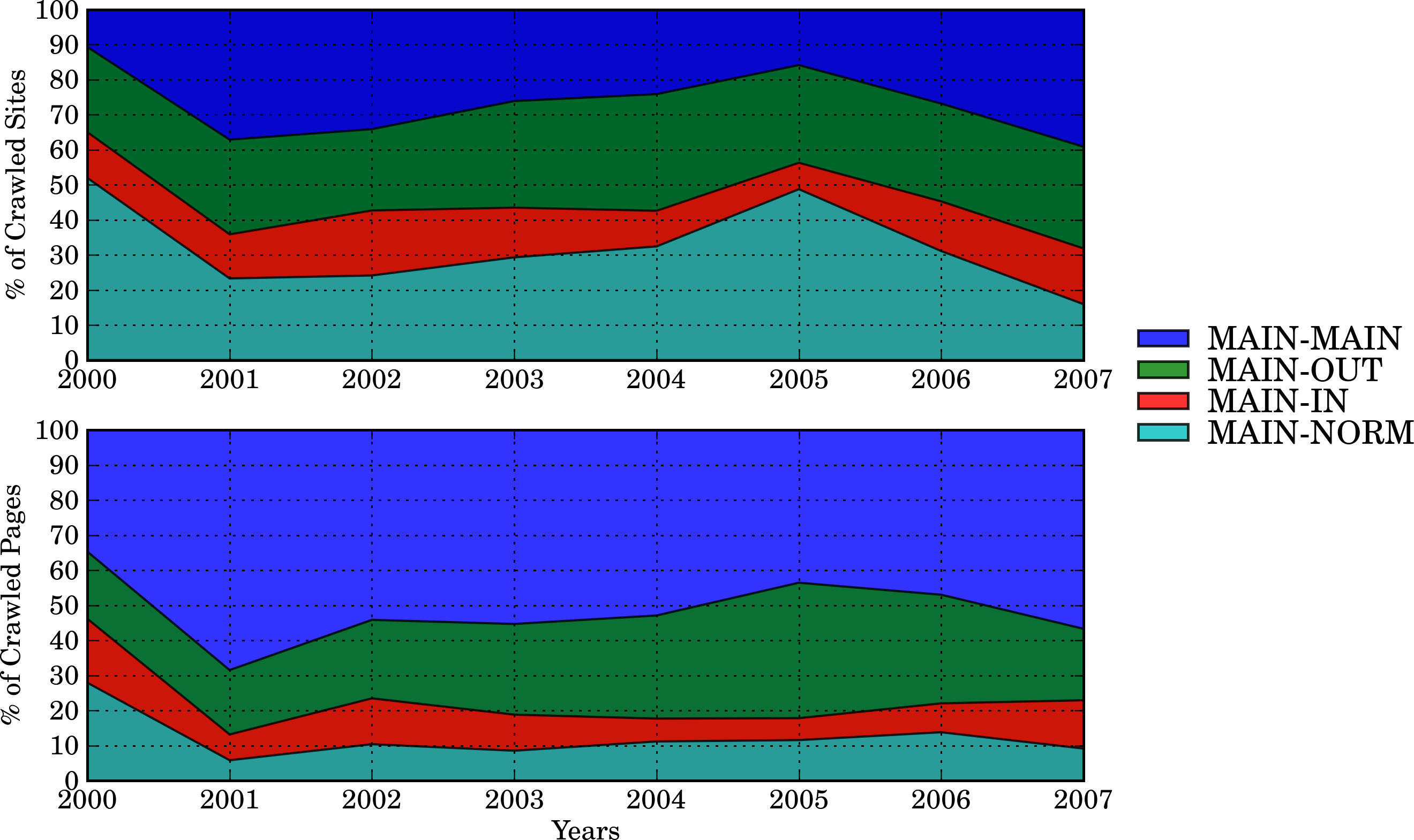}
 \caption{Evolution of the subcomponents of the MAIN component in terms of sites (top) and pages (bottom).}
 \label{fig:structure_evolution_main}
\end{figure}

Figure \ref{fig:structure_evolution} shows a graphical representation of the evolution of each component across the years. It becomes clear that, while MAIN is a small component in terms of the number of sites, it is the bigger component in terms of documents. ISLANDS is the component with most sites in almost all years, but it is not as big in terms of pages. Every year, approximately $50\%$ of sites in ISLANDS are sites with only one crawled page. Also, Figure \ref{fig:structure_evolution_main} shows the evolution of the MAIN subcomponents. In terms of pages they have a stable size, although in absolute size they are constantly growing. In terms of pages, MAIN-MAIN has shown an average near $50\%$ of the pages in MAIN. 

\subsection{Website migration}

The size of the components of the Web Structure change because of migrations from one component to another. A migration is defined as follows: if a site $S$ in some year is found in component $A$ and in the next year is found in component $B$ ($A \neq B$), we say that $S$ migrated from $A$ to $B$. 


Considering transitions between the defined components, we added special migrations: when a site is crawled for the first time, we add a transition from \textit{NEW} to the corresponding component for the site; when a site could not be crawled, we add a migration to a \textit{UNKNOWN} state; and when a site dies we add a migration to a \textit{DEAD} state. We calculated the aggregated migrations for all years and sorted them in descending order. The 10 most common migrations are shown in Table \ref{tab:transiciones_agregadas}. The most common is NEW-ISLAND, probably because it is natural for a new site to be disconnected from the others. Second is ISLAND-UNKNOWN and third ISLAND-DEAD, showing that a disconnected site in the hostgraph probably will disappear in the following years.

\begin{table}[hbt]
\begin{center}
\footnotesize
\caption{Total sorted percentage of migrations between components. The 10 most common migrations are shown.}
\begin{tabular}{lr}
\hline Transition & Percent \\
\hline NEW--ISLAND & $23,48$ \\
NEW--UNKNOWN & $5,69$ \\
ISLAND--UNKNOWN & $5,53$ \\
ISLAND--DEAD & $5,42$ \\
NEW--OUT & $5,22$ \\
UNKNOWN--DEAD & $4,32$ \\
NEW--MAIN & $3,96$ \\
UNKNOWN--ISLAND & $3,95$ \\
NEW--IN & $3,56$ \\
OUT--ISLAND & $3,07$ \\
\hline 
\end{tabular}
\label{tab:transiciones_agregadas}
\end{center}
\end{table}

\begin{table*}[bt]
\begin{center}
\footnotesize
\caption{Accumulated transitions per component.}
\begin{tabular}{l|ccccccccc}
\hline & MAIN & OUT & IN & ISLAND & TUNNEL & TIN & TOUT & UNKNOWN & DEAD \\
\hline MAIN & $43,44$ & $16,58$ & $6,81$ & $4,64$ & $0,40$ & $1,15$ & $1,27$ & $8,20$ & $17,53$ \\
OUT & $10,05$ & $42,42$ & $1,96$ & $17,69$ & $0,33$ & $3,17$ & $1,48$ & $10,71$ & $12,19$ \\
IN & $9,76$ & $5,21$ & $24,83$ & $19,60$ & $0,23$ & $1,23$ & $5,14$ & $13,64$ & $20,37$ \\
ISLAND & $1,49$ & $7,22$ & $3,37$ & $49,82$ & $0,09$ & $1,44$ & $3,06$ & $16,91$ & $16,57$ \\
TUNNEL & $12,51$ & $19,33$ & $5,84$ & $24,04$ & $2,98$ & $4,40$ & $9,82$ & $10,81$ & $10,24$ \\
TIN & $4,99$ & $22,71$ & $3,32$ & $31,28$ & $0,49$ & $9,18$ & $1,66$ & $13,42$ & $12,96$ \\
TOUT & $3,41$ & $8,02$ & $8,37$ & $32,46$ & $0,46$ & $1,09$ & $16,90$ & $15,12$ & $14,19$ \\
\hline UNKNOWN & $6,71$ & $11,74$ & $5,09$ & $32,46$ & $0,18$ & $1,55$ & $2,84$ & $3,99$ & $35,48$ \\
DEAD & $0,06$ & $0,12$ & $0,11$ & $0,85$ & $0$ & $0,02$ & $0,09$ & $3,19$ & $95,55$ \\
NEW & $8,90$ & $11,72$ & $8,02$ & $52,71$ & $0,17$ & $2,20$ & $3,49$ & $12,77$ & $0$ \\
\hline
\end{tabular}
\label{tab:transiciones_totales}
\end{center}
\end{table*}

Table \ref{tab:transiciones_totales} shows the aggregated transitions per component (including special migrations). Each row contains the corresponding percent of transitions. It is interesting to note that $43,42\%$ of sites in MAIN stay in MAIN from one year to another, a $42,40\%$ of sites in OUT stay in OUT, and $49,82\%$ of sites in ISLAND stay in ISLAND. A considerable percent of sites die each year, and this percent is similar across components. Another interesting point is the presence of transitions NEW-MAIN: if a new site is probably disconnected from the others, why there are new sites directly in MAIN and OUT? Considering the time passed between crawls it is possible for a new site to be linked and also give links in that period. A new site can easily be inside IN, because all it has to do is to put a link to a site in MAIN, but this doesn't happen so often: $8,02\%$ of the NEW transitions are NEW-IN. Although getting linked from other sites is harder, $8,9\%$ of the NEW transitions are NEW-MAIN and a $11,72\%$ are NEW-OUT. Of course, it is still more probably to become a new ISLAND, with $52,71\%$ of the NEW transitions. 

\section{Concluding remarks}
\label{sec:conclusions}

In this study we covered more than $50\%$ of the lifetime of the Chilean Web, which is similar to the time span of the whole Web in most countries, as it became popular in 1993. Although our results depend on the crawler used, we start from the complete list of \texttt{.cl} domains, which cannot be done in most countries. So the starting seed set is quite complete, and hence our crawls are more stable than others.

Our results confirm the exponential growth for the number of sites ($f_{n} = 1,32 f_{n-1}$), the number of pages ($f_{n} = 1,46 f_{n-1}$) and the content size ($f_{n} = 1,55 f_{n-1}$). Notice that the constant grows as we go one level down, a fact that can be explained by the underlying powerlaws involved.

The growth for sites is lower than the one found in \cite{dynamics-web-structure}, and matches the prediction done at that time, as there was a local saturation of domain names. This trend may change with the recent generic domain names approved by ICANN and more so with the later almost liberalization of names.
  
Even under chaotic circumstances, the Chilean Web have some characteristics of the Global Web, as shown by the approximations of powerlaws for most of the measures taken. Inside this chaos, we found that the government and some universities have the most stable sites in terms of number of documents, site content in megabytes and in-degree. 

Future work should include further study of domains, as they are more stable and meaningful. We also would like to compare the growth of the Chilean Web with the growth of the Internet users in Chile. Finally, with the new data we would like to study how the new social platforms are changing the structure of the Web. 

\textbf{Acknowledgments:} Thanks to NIC Chile for providing the \texttt{.cl} domain data, as well as the support of Millennium Nucleus Grant P04-067-F from Mideplan, Chile. Also, we would like to thank the anonymous reviewers for their useful comments.

\bibliographystyle{plain}
\bibliography{evochiweb}

\end{document}